%% file: main.tex
\documentclass[10pt, conference, letterpaper, table]{IEEEtran}

\IEEEoverridecommandlockouts
\usepackage{graphicx}
\usepackage{subcaption}
\usepackage{pgf} 

\ifCLASSOPTIONcompsoc
 \usepackage[caption=false, font=normalsize, labelfont=sf, textfont=sf]{subfig}
\else
\usepackage[caption=false, font=footnotesize]{subfig}
\fi
\usepackage{cite}
\usepackage{url}
\usepackage{mdwmath} 
\usepackage{soul}
\usepackage{ifthen}
\usepackage{amsfonts,amsmath,amssymb,amsxtra,amsbsy,floatflt}
\usepackage{indentfirst} 
\usepackage{tabularx}
\usepackage{booktabs}
\usepackage{multirow}
\usepackage{nth}
\usepackage{comment}
\usepackage{epstopdf}
\usepackage{hhline}
\usepackage{siunitx}
\usepackage{booktabs}
\usepackage[bottom]{footmisc}
\usepackage[acronym]{glossaries}
\usepackage{upgreek}
\usepackage{algorithm}
\usepackage{algorithmicx,algpseudocode}
\usepackage{bm}
\usepackage[USenglish,UKenglish]{babel}

\input{Definitions.tex}

\input{Acro.tex}

\title{Curved Apertures for Customized Wave Trajectories: Beyond Flat Aperture Limitations}

\author{Joan Mart\'inez Canals,
Francesco Devoti,~\IEEEmembership{Member,~IEEE,}
Vincenzo~Sciancalepore,~\IEEEmembership{Senior Member,~IEEE,}\\
Marco Di Renzo,~\IEEEmembership{Fellow,~IEEE,}
and Xavier~Costa-P\'erez,~\IEEEmembership{Senior Member,~IEEE}\thanks{\textit{J. Mart\'inez-Canals is with i2cat. F. Devoti and V. Sciancalepore are with NEC Laboratories Europe. M. Di Renzo is with Universit\'e Paris-Saclay, CNRS, CentraleSup\'elec, Laboratoire des Signaux et Syst\`emes (marco.di-renzo@universite-paris-saclay.fr), and with King's College London, Centre for Telecommunications Research -- Department of Engineering (marco.di\_renzo@kcl.ac.uk). X. Costa-P\'erez is with i2cat, ICREA, and NEC Laboratories Europe.
\newline This work was supported by the European Commission’s Horizon Europe Programme under the SNS JU INSTINCT project (Grant Agreement 101139161), by the CFIS Mobility Program funded by Fundació Privada Mir-Puig, CFIS partners, and donors of the crowdfunding program, and by the Engineering and Physical Sciences Research Council (EPSRC), part of UK Research and Innovation, and the UK Department of Science, Innovation and Technology (grant reference EP/X040569/1).
\newline Email of the corresponding author: francesco.devoti@neclab.eu}.}}

\begin{document}

\maketitle

\begin{abstract}
Beam shaping techniques enable tailored beam trajectories, offering unprecedented connectivity opportunities in wireless communications. Current approaches rely on flat apertures, which limit trajectory flexibility due to inherent geometric constraints. To overcome such restrictions, we propose adopting curved apertures as a more versatile alternative for beam shaping. We introduce a novel formulation for wave trajectory engineering compatible with arbitrarily shaped apertures. Theoretical and numerical analyses demonstrate that curved apertures offer improved control over wave propagation, are more resilient to phase control constraints, and achieve higher power density across a wider portion of the desired beam trajectory than flat apertures.
\end{abstract}

\begin{IEEEkeywords}
Curved beams, Self-accelerating beams, Caustic beams, Airy beams.
\end{IEEEkeywords}

\glsresetall

\section{Introduction}
\label{sec:intro}
The design of wave trajectories has gained significant attention in recent years due to their applications in optics, acoustics, and electromagnetism~\cite{efremidis2019airy, zhao2014delivering, guerboukha2024curving}. Airy beams exhibiting a parabolic trajectory and non-diffracting properties were initially introduced as solutions of the paraxial wave equation~\cite{efremidis2019airy}, stimulating interest in beam engineering and inspiring further formulations~\cite{kaminer2012nondiffracting, zhang2012nonparaxial}. Such beams were found to have an underlying caustic structure~\cite{berry2017stable}, where the envelope of propagating rays is tangent to a curve or surface, namely \emph{the caustic}, thereby creating a region of concentrated energy. Beams following a desired caustic can be engineered directly from geometrical considerations~\cite{trajectories}, pioneering the concept of caustic beams.
Trajectory engineering techniques open several possibilities, which can bypass the capabilities of current technologies, such as \acrlongpl{ris} in overcoming obstacles~\cite{encinas2024cost}: Instead of depending on external electromagnetic structures to programmatically redirect waves~\cite{pan2021reconfigurable}, a tailored-designed wavefront can navigate complex environments more efficiently~\cite{guerboukha2024curving}. This technology shift enhances the adaptability of wave-based applications and reduces reliance on supplementary control mechanisms, enabling more streamlined and robust implementations in \acrlong{bte}.

Traditionally, beam shaping relies on planar apertures and phase modulation to customize wave propagation. However, this conventional formulation faces several inherent limitations. First, flat apertures are not universally suitable for all applications, as their geometric constraints may seriously limit the achievable trajectories~\cite{droulias2024bending}. Second, the generation of structured wavefronts through planar surfaces is typically constrained to simple refraction and reflection angles (e.g., $90$-degree bends)~\cite{droulias2024bending}. This limitation reduces the flexibility required for complex beam path designs~\cite{mursia2024t3dris}.

To overcome such constraints, in this letter, we argue that curved apertures may represent a powerful alternative: By leveraging their intrinsic ability to modulate the phase more effectively, curved apertures enable smoother and more customized beam trajectories. One key advantage of this solution is the ability to relax phase fluctuations at the aperture, which facilitates the practical engineering of complex wave patterns that would otherwise be arduous using traditional flat surfaces. Additionally, curved surfaces provide more robust control over the parameters shaping the beam trajectory, making it feasible to engineer continuous and non-trivial paths without excessive reliance on external phase-correcting technologies.

This paper fully explores the potential of caustic beams shaped by curved apertures, shedding light on their advantages in tailoring customized trajectories. We introduce a novel framework for trajectory engineering applicable to apertures of arbitrary geometry. Through theoretical analysis and numerical simulations, we unveil fundamental limitations inherent in flat-aperture configurations and demonstrate that employing curved apertures enhances the control of wave propagation, surpassing the constraints imposed by conventional flat designs.

\section{Caustic-beam Formulation}
\label{sec:problem_formulation}
We provide the traditional formulation tailored to flat apertures, followed by a generic formulation for generating caustic beams with arbitrarily shaped trajectories.

\subsection{Traditional Formulation}
\begin{figure}
\centering
 \begin{subfigure}[b]{0.45\linewidth}
 \includegraphics[width=0.9\linewidth]{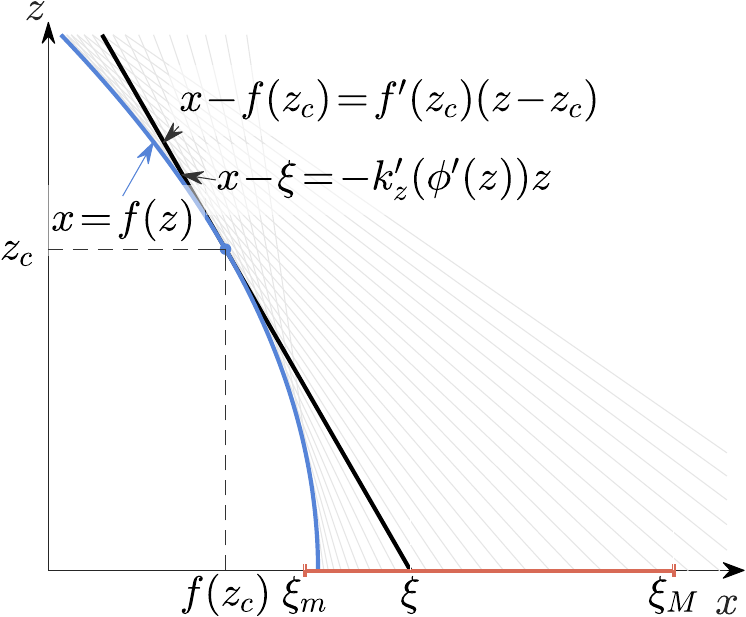}
 \caption{Flat aperture} \label{fig:geometry_flat}
 \end{subfigure}
 \hfill
 \begin{subfigure}[b]{0.49\linewidth} \includegraphics[width=0.9\linewidth]{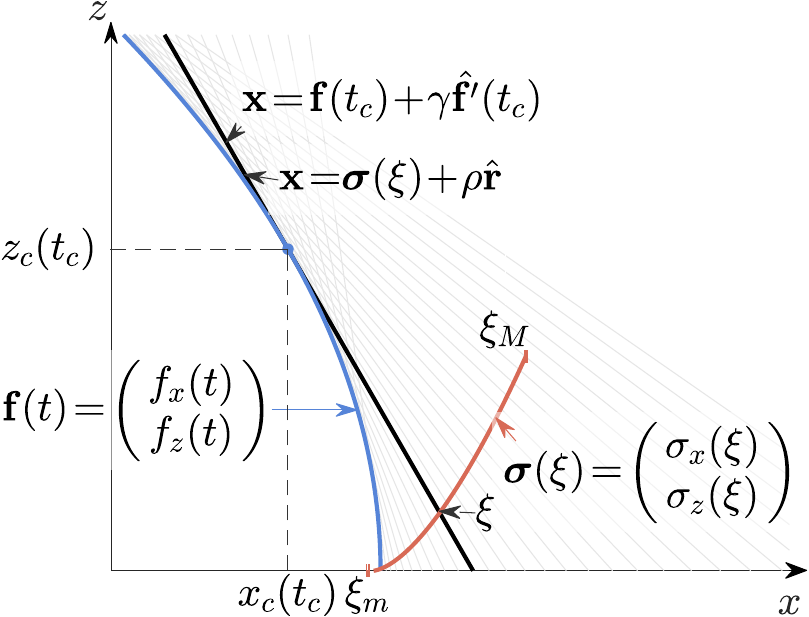}
 \caption{Arbitrary aperture} \label{fig:geometry_curve}
 \end{subfigure}
 \caption{Geometry considered in the case of flat apertures and the general case of arbitrary apertures.}
\label{fig:geometry}
\end{figure}

We consider a \acrlong{2d} geometry, the $xz$-plane in particular, with a continuous aperture on the $x$-axis, as illustrated in Fig.~\ref{fig:geometry_flat}, i.e., at $z \!=\! 0$, and the $z$-axis parallel to the direction of propagation. We consider a $y$-polarized electric field with amplitude $\psi(x,z)$, which is a scalar function that depends on the transverse coordinate $x$ and the propagation coordinate $z$, i.e., $\mE \!=\! \hat{\vy}\psi(x,z)e^{\!-i\omega t}$. The field $\psi(x,z)$ results from the propagation of the field $\psi_0(\xi) \!=\! A(\xi)e^{i\phi(\xi)}$ imposed at the aperture,\footnote{We focus on geometric aspects of \acrlong{bte}, by neglecting field polarization and mode matching considerations and treating the field as a scalar quantity. Consequently, we use the terms ``field" and ``field amplitude" interchangeably.\label{foot:field}} where $A(\xi)$ and $\phi(\xi)$ are the amplitude and phase of $\psi_0(\xi)$ at the aperture point $\xi$, respectively. Let us assume $kr \gg 1$, where $k$ is the wave number, and $r \!=\! \sqrt{(x\!-\!\xi)^2\!+\!z^2}$. By using the Rayleigh-Sommerfeld formulation, $\psi(x,z)$ can be formulated as follows~\cite{trajectories}:
\begin{equation}
\psi(x,z)\!=\!2\int_{\xi_m}^{\xi_M} \psi_0(\xi) \frac{\partial G(x,z;\xi)}{\partial z}\,d\xi, \label{eq:field}
\end{equation}
where $G(x,z;\xi) \!=\! \!\frac{-i}{4}H_0^{(1)}\!(kr)$ is the Green function. $H_0^{(1)}\!(\cdot)$ is the first-kind, first-order Hankel function, with $\frac{\partial G(x,z;\xi)}{\partial z} \!=\! \frac{i}{4}\frac{kz}{r}H_1^{(1)}(kr)$. Also, $\xi_m$ and $\xi_M$ are the aperture boundaries.

We aim to find $\psi_0(\xi)$ generating a caustic beam along the desired trajectory $x \!=\! f(z)$. Caustic beams are formed by the constructive interference of propagating waves along specific paths (rays), leading to high-intensity regions with sharp features~\cite{chen2023generation}. The beam trajectory is controlled by $\phi(\xi)$, while $A(\xi)$ controls the intensity profile and can be set arbitrarily~\cite{trajectories}. Focusing on the design of $\phi(\xi)$, we derive the ray equation describing the path along which the phase remains stationary. To do so, we leverage the stationary phase approximation method to identify points where the phase of the oscillating integral in \eqref{eq:field} is stationary, allowing us to isolate and calculate the dominant contributions effectively. Specifically, considering the condition $kr\!\gg\!1$, we leverage the approximation $H_1^{(1)}\!(kr) \!\approx\! \sqrt{\!\frac{2}{\pi k r}}e^{i(k r - \frac{3\pi}{4})}$, and apply the stationary phase condition $\frac{\partial}{\partial \xi} \! \!\left( k r \!-\! \frac{3\pi}{4} \!+\! \phi(\xi) \right) \!=\! 0$, which yields the ray equation $x \!=\! \frac{r}{k}\phi'(\xi) \!+\! \xi$.
By considering $r\!=\!\sqrt{(x\!-\!\xi)^2\!+\!z^2}$, $k\!=\!\sqrt{k_x^2\!+\!k_z^2}$, and $k_x\!=\!\phi'(\xi)$,\footnote{Note that the wave vector is $\vk\!=\!\nabla \phi(x,y,z)$, so $k_x\!=\!\phi'(\xi)$.} and substituting them in the ray equation we obtain
\begin{equation}
    x \!=\! \xi \mp k'_z(\phi'(\xi))z \rightarrow x \!=\! \xi \!-\!k'_z(\phi'(\xi))z, \label{eq:ray_3}
\end{equation}
where $k_z(k_x)\!=\!\sqrt{k^2\!-\!k^2_x}$ is the dispersion relation obtained by selecting the solution in the first quadrant considering $k_x^2\!+\!k_z^2\!=\!k^2$ \cite{trajectories}, from which $k_z(k_x) \!=\! \sqrt{k^2\!-\!k_x^2} \!=\! k_z(\phi'(\xi)) \!=\! \sqrt{k^2\!-\!\phi'^2(\xi)}$, and $k_z'(\phi'(\xi))\!=\!\frac{d\,k_z(\phi'(\xi))}{d\,\phi'(\xi)}$. Note that the sign of $x\!-\!\xi$ depends on the bending direction of $f(z)$. In Fig.~\ref{fig:geometry_flat}, e.g., as long as the condition $\xi \!\geq\! x$ holds, we have $ x\!-\!\xi \!\leq\! 0$.

The line tangent to the trajectory and crossing the point $(x_c,z_c)\!=\!(f(z_c),z_c)$ is described as
\begin{equation}
 x\!-\!f(z_c)\!=\!f'(z_c)(z\!-\!z_c), \label{eq:tangent_line}
\end{equation}
where $f'(z_c) \!=\! \frac{d\,f(z_c)}{d\,z_c}$. The desired phase profile is obtained by enforcing that \eqref{eq:ray_3} satisfies \eqref{eq:tangent_line} at $z\!=\!z_c$, i.e., the ray is tangent to the trajectory. In other words, we impose the condition
\begin{equation}
 f(z_c)\!=\! - k_{z}'(\phi'(\xi))z_c \!+\! \xi. \label{eq:caustic_condition}
\end{equation}
By derivation of \eqref{eq:caustic_condition} with respect to $z_c$, we obtain $k_{z}'(\phi'(\xi)) \!=\! \!-\!f'(z_c)$. Therefore, the relation between $z_c(\xi)$ and $\phi({\xi})$ is
\begin{equation}
 k'_z(\phi'(\xi))\!=\!\frac{-\phi'(\xi)}{\sqrt{k^2\!-\!\phi'^2(\xi)}}\!=\!-\!f'(z_c(\xi)).
\end{equation}
With the aid of some manipulations, we obtain 
\begin{equation}
\phi'(\xi)\!=\!\frac{kf'(z_c(\xi))}{\sqrt{1\!+\!f'^2(z_c(\xi))}}, \label{eq:phase_flat}
\end{equation}
and, by direct integration, $\phi(\xi) \!=\! \int \phi'(\xi) \,d\xi$. Eq. \eqref{eq:phase_flat} is general and is derived without relying on simplifying assumptions on propagation conditions such as the paraxial regime. Finally, $z_c(\xi)$ is obtained from \eqref{eq:tangent_line} by imposing $x\!=\!\xi$ at the interception of the tangent line with the aperture ($z\!=\!0$), i.e., $\xi\!=\!f(z_c)\!-\!f'(z_c)z_c\!=\!g(z_c)$, from which we obtain $z_c(\xi)\!=\!g^{-1}(\xi)$.\footnote{Note that solving $g^{-1}(\xi)$ in closed-form requires knowledge of $f(z_c)$.}

\subsection{General Formulation}
The state-of-the-art literature on caustic beams is focused on particular cases, such as considering a straight aperture (e.g., lying on one of the axes) or having a bijective function describing the desired caustic trajectory (e.g., $f(z)\!=\!x$). In this section, we relax those assumptions and consider an arbitrary choice for the trajectory $\vf(t) \!=\! (f_x(t),f_z(t))^{\tran}$, and the aperture $\sigmab(\xi) \!=\! (\sigma_x(\xi),\sigma_z(\xi))^{\tran}$, as illustrated in Fig.~\ref{fig:geometry_curve}. Note that the straight aperture on the $x$-axis ($\sigmab(\xi) \!=\! (\xi,0)^{\tran}$) and the bijective trajectory ($\vf(t) \!=\! (f_x(t),t)^{\tran}$, i.e., $x_c \!=\! f(z_c)$) are both special cases of $\sigmab(\xi)$ and $\vf(t)$, as shown in Fig.~\ref{fig:geometry_flat}.

In this case, assuming $kr \gg 1$ and near field conditions, the Rayleigh-Sommerfeld formulation can be applied as a line integral parametrized along the aperture~\cite{mansuripur2020tutorial}
\begin{equation}
 \psi(\vx) \!=\! 2\int_{\xi_m}^{\xi_M}A(\xi)e^{i\phi(\xi)} \partial_{\hat{\vn}(\xi)} G(\vx;\xi)\| \sigmab'(\xi)\|\, d\xi,
\label{eq:curvedIntegral}
\end{equation}
where $\vx\!=\!(x,z)^{\tran}$ is a point in the space, $\partial_{\hat{\vn}(\xi)}G(\vx;\xi)\!=\!\nabla G(\vx;\xi)\hat{\vn}(\xi)$ denotes the directional derivative of the Green function along the unitary vector $\hat{\vn}(\xi)$ normal to the aperture at the coordinate $\xi$,\footnote{The hat symbol ( $\hat{}$ ) denotes both unit vectors and vector normalization.} and $\sigmab'(\xi)\!=\!(\sigma_x'(\xi),\sigma_z'(\xi))^{\tran}$.  $\|\sigmab'(\xi)\|$ is the norm of the parametrization of the line integral, and $r(\xi) \!=\!\|\vr(\xi)\| \!=\! \|\vx\!-\!\sigmab(\xi)\|$. The derivative $\partial_{\hat{\vn}(\xi)}G(\vx;\xi)$ is
\begin{equation}
\partial_{\vn(\xi)}G(\vx;\xi)
 \!=\! \frac{i k}{4 r}H_1^{(1)}(kr(\xi)) \vr(\xi)^{\tran}\hat{\vn}(\xi).
\label{eq:normalDerivative}
\end{equation}
By considering $kr\!\gg\!1$, we apply the stationary phase method to \eqref{eq:curvedIntegral}, and substitute \eqref{eq:normalDerivative} in \eqref{eq:curvedIntegral}, obtaining
\begin{equation}
 \psi(\vx)\!\approx\! \int_{\xi_m}^{\xi_M}M(\xi)e^{i(\phi(\xi)+kr(\xi)-\frac{3}{4}\pi)}\, d\xi,
 \label{eq:asymptoticCurvedIntegral}
\end{equation}
with $M(\xi)\!=\!\frac{i}{2}A(\xi)\vr(\xi)^{\tran}\vn(\xi)\,\|\sigmab'(\xi)\,\|\sqrt{\frac{2}{\pi kr(\xi)}}$. The corresponding ray equation is $\frac{d}{d\xi}\big(\phi(\xi)\!+\!kr(\xi)\!-\!\frac{3}{4}\pi\big)\!=\! \phi'(\xi)\!-\!\hat{\vr}(\xi)^{\tran}\sigmab'(\xi)k\!=\!0$. From it, we obtain
\begin{equation}
\phi'(\xi)\!=\!k\hat{\vr}^{\tran}\sigmab'(\xi), \label{eq:phase_profile_vec_1}
\end{equation}
which expresses $\phi'(\xi)$ as a function of the ray direction $\hat{\vr}$ and the aperture shape. The corresponding ray is described by  
\begin{align}
 \vx \!=\! \sigmab(\xi) \!+\! \rho \hat{\vr},\,\,\, \rho \in \Real.\label{eq:ray_equation_free}
\end{align}

Let us consider a given desired caustic trajectory $\vf(t)$. The tangent line passing through the point $\vf(t_c)$ is described by
\begin{equation}
 \vx\!=\!\vf(t_c) \!+\! \gamma \hat{\vf}'(t_c), \,\,\, \gamma \in \Real, \label{eq:tangent_line_free}
\end{equation}
with $\vf'(t) \!=\! \frac{d\,\vf(t)}{d\,t}$. Since a straight line tangent to the trajectory is a ray equation, we can derive the desired phase profile by imposing that \eqref{eq:ray_equation_free} satisfies \eqref{eq:tangent_line_free} in $t\!=\!t_c$. Hence, by aligning the tangent line and the ray equation, i.e., $\hat{\vf}'(t_c)\!=\!\hat{\vr}$, and plugging this condition in \eqref{eq:phase_profile_vec_1}, we obtain
\begin{equation}
\phi'(\xi)\!=\!k\hat{\vf}'(t_c(\xi))^{\tran}\sigmab'(\xi), \label{eq:phase_profile_free}
\end{equation}
which gives us the desired phase profile by direct integration
\begin{equation}
\phi(\xi)\!=\!\int_{\xi_m}^{\xi} k\hat{\vf}'(t_c(\xi))^{\tran}\sigmab'(\xi) d\xi, \label{eq:phase_profile_free_int}
\end{equation}
where $t_c(\xi)$ returns the value of $t_c$ such that the tangent line passing through $\vf(t_c)$ intercepts with the aperture in $\sigmab(\xi)$.

From \eqref{eq:phase_profile_free_int}, the desired phase is the integral of the tangent vector to the aperture $\sigmab'(\xi)$ projected onto the direction of the line to the trajectory $\hat{\vf}'(t_c(\xi))$, scaled by $k$. This represents a local beamforming toward the tangent direction of the trajectory at the aperture point $\xi$.

\section{Theoretical limits of straight and curved apertures in generating beam trajectories}
\label{sec:limits}

As discussed in Section~\ref{sec:problem_formulation}, the caustic trajectory strictly depends on the phase variation $\phi'(\xi)$ across the aperture. However, in practical scenarios, such phase variation can be constrained by factors related to implementation complexity and cost considerations~\cite{trastoy2001phase}. Smooth phase variations across the aperture are advisable to prevent abrupt changes, which may otherwise result in unwanted side lobes or distortions of the radiation pattern~\cite{mazzinghi2018enhanced}. Furthermore, in the context of antenna arrays, the phase distribution must be sampled at intervals compliant with the Nyquist criterion to avoid spatial aliasing~\cite{droulias2024bending}. Therefore, the spacing between array elements imposes an inherent limitation on the maximum achievable phase variation. Thus, it becomes necessary to define a practical boundary $d\Phi$ on the phase variation $\phi'(\xi)$, such that $|\phi'(\xi)| \leq d\Phi$, for all $\xi \in \mathcal{D}$. Then, for any given trajectory $\vf(t)$ with $|\phi'(\xi)| \!\leq\! d\Phi $ enforced,\footnote{This condition is on the derivative since we consider continuous apertures, but it can be readily extended to discrete apertures, i.e., antenna arrays.} the zones of the aperture where $|\phi'(\xi)| \!>\! d\Phi $ (if exist) will not constructively contribute in generating the beam, leading to energy loss along the trajectory. Note that realizing such controlled phase profiles requires reconfigurable elements (e.g., using PIN diodes, varactors, or phase-changing materials), which affect system complexity, cost, and calibration. Curved apertures, in particular, pose additional challenges in maintaining precise phase alignment. Such aspects are critical in transitioning from conceptual models to deployable systems and will be further addressed in future experimental work. Below, we analyze the theoretical limits of $|\phi'(\xi)|$ for straight and curved apertures.

\subsection{Straight Aperture}
\begin{figure}
 \centering
 \begin{subfigure}[b]{0.24\textwidth}
 \centering
\includegraphics[width=0.9\textwidth]{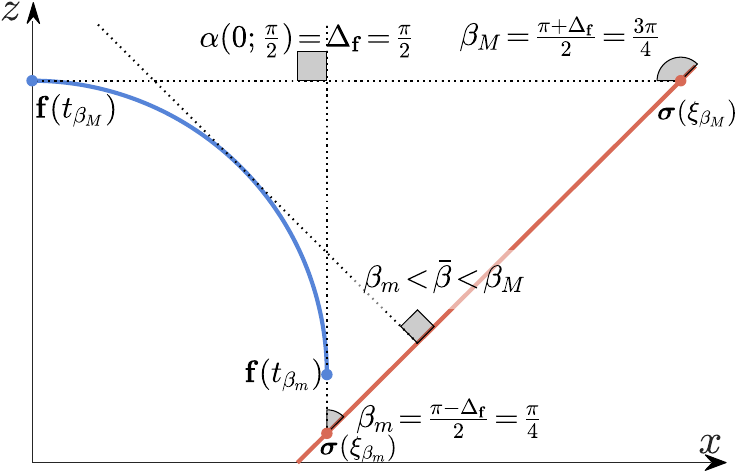}
 \caption{Quantities $\beta(\cdot)$, and $\alpha(\cdot)$.}
 \label{fig:anglesAlpha}
 \end{subfigure}
 \hfill
 \begin{subfigure}[b]{0.24\textwidth}
 \centering
 \includegraphics[width=0.9\textwidth]{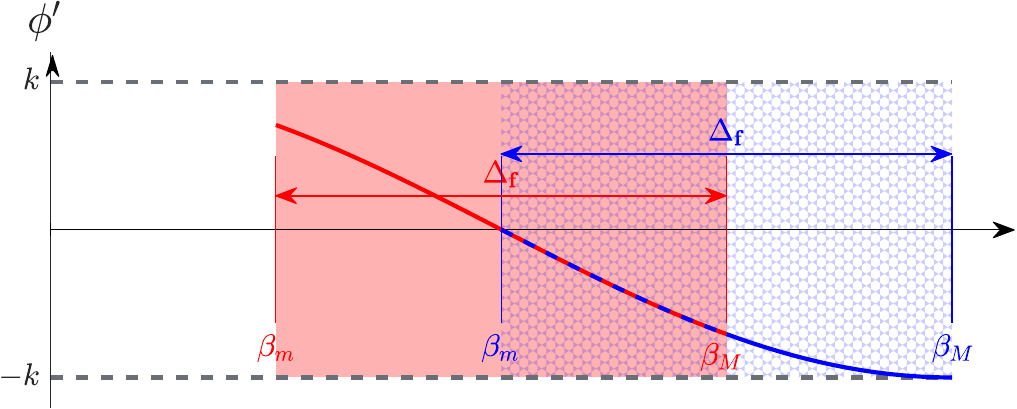}
 \hfill
 \caption{
 Interval $\Delta_{\vf}$.}
 \label{fig:betaD_explained}
 \end{subfigure}
 \caption{Quantities $\beta(\cdot)$, $\alpha(\cdot)$, and $\Delta_{\vf}$ involved in generating a circular caustic (blue) with a straight aperture (red). Alongside, illustration of the interval $\beta(\mathcal{D})$, with span $\Delta_{\vf}$ imposed by the trajectory, and shift $\beta_m$ depending on the aperture.} \label{fig:angles_and_beta}\vspace{-5mm}
\end{figure}
Let us consider a straight aperture parametrized as $\sigmab(\xi)\!=\!\sigmab_0\!+\!\xi\hat{\vv}, \text{ } \xi \in \mathcal{D}$, with $\mathcal{D}\!=\![\xi_m, \xi_M]$ being the domain of the aperture, and an arbitrary trajectory, $\vf(t)$. Let us also assume $\mathcal{D}$ and $\vf(t)$ such that all tangent lines of $\vf(t)$ intersect with the aperture, i.e., $t_c(\xi)$ is defined $\forall \xi \in \mathcal{D}$.  From \eqref{eq:phase_profile_free}, we have
\begin{equation}
 |\phi'(\xi)|\!=\!k|\cos(\beta(\xi))|,
 \label{eq:phase_derivative_cosine}
\end{equation}
where $\beta(\xi)\,\,\!=\!\,\,\angle(\hat{\vv},\vf'(t_c(\xi)))$ is the angle between $\hat{\vv}$ and $\vf'(t_c(\xi))$. We refer the reader to Fig.~\ref{fig:angles_and_beta} for a clear visualization of the quantities involved.

Let us introduce the aperture points $\xi_{\beta_m}$, and $\xi_{\beta_M}$ corresponding to the minimum and maximum values of $\beta(\cdot)$, i.e., 
\begin{equation}
\xi_{\beta_m} \!=\! \argmin\limits_{\xi \in \mathcal{D}}\beta(\xi)\text{, and } \xi_{\beta_M}\!=\! \argmax\limits_{\xi \in \mathcal{D}}\beta(\xi),
\end{equation} 
and the trajectory points $t_{\beta_m}\!=\!t_c(\xi_{\beta_m})$ and $t_{\beta_M}\!=\!t_c(\xi_{\beta_M})$. Then, we have the values $\beta_m = \beta(\xi_{\beta_m})$, and $\beta_M = \beta(\xi_{\beta_M})$. Accordingly, $\beta_m \!\leq\! \beta(\xi) \!\leq\! \beta_M \,\, \forall \xi \in \mathcal{D}.$ Hence, the values of $\beta(\xi)$ are in the closed interval $\beta(\mathcal{D})\!=\![\beta(\xi_{\beta_m}),\beta(\xi_{\beta_M})]$.\footnote{Since $\beta$ is continuous and $\mathcal{D}$ is compact and connected in $\mathbb{R}$, $\beta(\mathcal{D})$ is a compact connected set of $\mathbb{R}$, and thus a closed interval.}
Since $\hat{\vv}$ is fixed, for any $(\xi_k,\xi_l)$ such that $\xi_k,\xi_l \in \mathcal{D}$, we have
\begin{equation}
\beta(\xi_l)\!=\!\beta(\xi_k)\!+\!\alpha(t_c(\xi_k),t_c(\xi_l)),
 \label{eq:beta_angle_progression}
\end{equation}
where $\alpha(t_k,t_l)\!=\!\angle(\vf'(t_c(\xi_k)),\vf'(t_c(\xi_l))$. From \eqref{eq:beta_angle_progression}, we get
\begin{equation}
\beta(\xi_{\beta_M})\!-\!\beta(\xi_{\beta_m})\!=\!\beta_M\!-\!\beta_m\!=\!\alpha(t_{\beta_m},t_{\beta_M})\!=\!\Delta_{\vf},
\end{equation}
hence $\beta(\mathcal{D})\!=\!\beta_m\!+\![0,\Delta_{\vf}]$. Accordingly, the span of $\beta(\mathcal{D})$, $\Delta_{\vf}$, is  constant  and independent  of $\hat{\vv}$, while  $\hat{\vv}$  is shifting $\beta(\mathcal{D})$ in

\noindent $\mathbb{R}$. Moreover, given $\beta(\mathcal{D})$, the minimum value $|\phi'|_m$ and maximum value $|\phi'|_M$ of the phase derivative are the following
\begin{equation}
|\phi'|_m \!=\!\! \min\limits_{\beta \in \beta(\mathcal{D})} \!k|\cos(\beta)|\text{, and } |\phi'|_M \!=\!\! \max\limits_{\beta \in \beta(\mathcal{D})} \!k|\cos(\beta)|.
\end{equation}

As a result, \textit{for straight apertures, the trajectory itself imposes variations of the phase derivative, while the aperture modulates the maximum and minimum values of these variations}. An example is provided in Fig.~\ref{fig:betaD_explained}.

\subsection{Best-case Scenario for Straight Apertures}
Now we consider the best-case scenario minimizing $|\phi'|_M$. From \eqref{eq:phase_derivative_cosine}, and if we restrict the values of $\beta(\cdot)$ to $(-\pi,\pi]$, $|\phi'|_M$ is achieved at $\xi$ such that $\beta(\xi)$ is the value furthest away from $\pm\frac{\pi}{2}$. In other words, $\xi\!=\!\argmax_{\xi\in \mathcal{D}}\left\{\left||\beta(\xi)|\!-\!\frac{\pi}{2}\right|\right\}$. Let us define $\xi_{\bar{\beta}}\!=\!\beta^{-1}\left(\beta_m\!+\!\frac{\Delta_{\vf}}{2}\right)$ as the value of $\xi$ corresponding to the value $\bar{\beta}$ of $\beta(\xi)$ at the middle of $\Delta_{\vf}$, which yields
\begin{equation}
|\phi'|_M \!=\! k \Big|\cos \Big(\frac{\pi}{2} \!+\! \min \Big\{ \Big||\bar{\beta}| \!-\! \frac{\pi}{2}\Big| \!+\! \frac{\Delta_{\vf}}{2},\frac{\pi}{2}\Big\}\Big) \Big|.
\label{eq:general_case_flat}
\end{equation}
From \eqref{eq:general_case_flat}, it is clear that, if $\Delta_{\vf}\!\geq\!\pi$, $|\phi'|_M\!=\!k|\cos(\pi)|\!=\!k$, for any $\hat{\vv}$. Whereas if $\Delta_{\vf}\!<\!\pi$, by choosing $\hat{\vv}$ such that $\bar{\beta}\!=\!\pm\frac{\pi}{2}$, the minimum possible value for $|\phi'|_M$ is
\begin{equation}
 |\phi'|_M \!=\! k\Big|\cos\Big(\frac{\pi\!+\!\Delta_{\vf}}{2}\Big)\Big|. \label{eq:best_case_flat}
\end{equation}
We want to stress that \eqref{eq:best_case_flat} gives us the best case when $\hat{\vv}$ minimizes $|\phi'|_M$, while the general case is reported in~\eqref{eq:general_case_flat}.

To better understand, let us consider the example of a circular trajectory of radius $R$ described as $\vf(t)\!=\!R(\cos(t/R),\sin(t/R))$; $t\in[0,\pi R / 2 ]$. In this case, $\alpha(0,t)\!=\!\frac{t}{R}$. Therefore, $\Delta_{\vf}\!=\!\pi/2$ and $\bar{t}\!=\!t_c^{-1}(\xi_{\bar{\beta}})\!=\!\pi R/4$ Since $\vf'(\bar{t})\!=\!\tfrac{1}{\sqrt{2}}(-1,1)^{\tran}$, $|\phi'|_M$ is minimized when $\hat{\vv}\!=\!\tfrac{1}{\sqrt{2}}(1,1)^{\tran}$, as it is perpendicular to $\hat{\vf}'(\bar{t})$. In this case, the aperture that minimizes $|\phi'|_M$ is $\sigmab(\xi) \!=\! \mathbf{\sigmab(\xi_m)}\!+\!\frac{\xi}{\sqrt{2}}(1,1)^{\tran}$, where $\sigmab(\xi_m)$ is an initial point for the aperture such that $t_c(\xi)$ is defined $\forall \xi \in \mathcal{D}$. From \eqref{eq:best_case_flat}, $|\phi'|_M\!=\!\frac{k}{\sqrt{2}}$.

Another limitation of the straight aperture arises when considering an arbitrary $\hat{\vv}$ with nonnegative components $v_x$ and $v_z$ and a trajectory with a single bending direction (e.g., a curve with a constant curvature sign), starting from an aperture extreme, i.e., $\xi(0)\!=\!\xi_m$ or $\xi_M$. If $\Delta_{\vf}\!\geq\!\pi$, the trajectory cannot be generated with the considered straight aperture. Indeed, for $\xi(0)\!=\!\xi_m$, the trajectory requires $\xi(t)$ and $t_c(\xi)$ to be increasing. Moreover, since $\hat{\vv}$ has nonnegative components, the trajectory's curvature must also be nonnegative, making $\alpha(0;\cdot)$ an increasing function. Therefore, the condition $\beta(\mathcal{D})\!=\!\beta(0)\!+\![0,\Delta_{\vf}]$ (see~\eqref{eq:beta_angle_progression}) implies the existence of a $t_k\!>\!0$ such that $\beta(t_k)\!=\!\pi$ or $2\pi$. In other words, at $\vf(t_k)$, the tangent line is parallel to the aperture, preventing generating the caustic. The same argument remains valid when considering the initial condition $\xi(0)\!=\!\xi_M$; however, in this case, both $\xi(t)$ and $t_c(\xi)$ become decreasing functions. Consequently, the curvature attains non-positive values, and the resulting angular range is given by $\beta(\mathcal{D})\!=\!\beta(0)\!+\![-\Delta_{\vf},0]$.
Likewise, it can be demonstrated that for trajectories with constant curvature that do not intersect the aperture, the condition $\Delta_{\vf}\!<\!\pi$ holds.
\begin{figure*}
 \centering
 \includegraphics[width=0.87\linewidth]{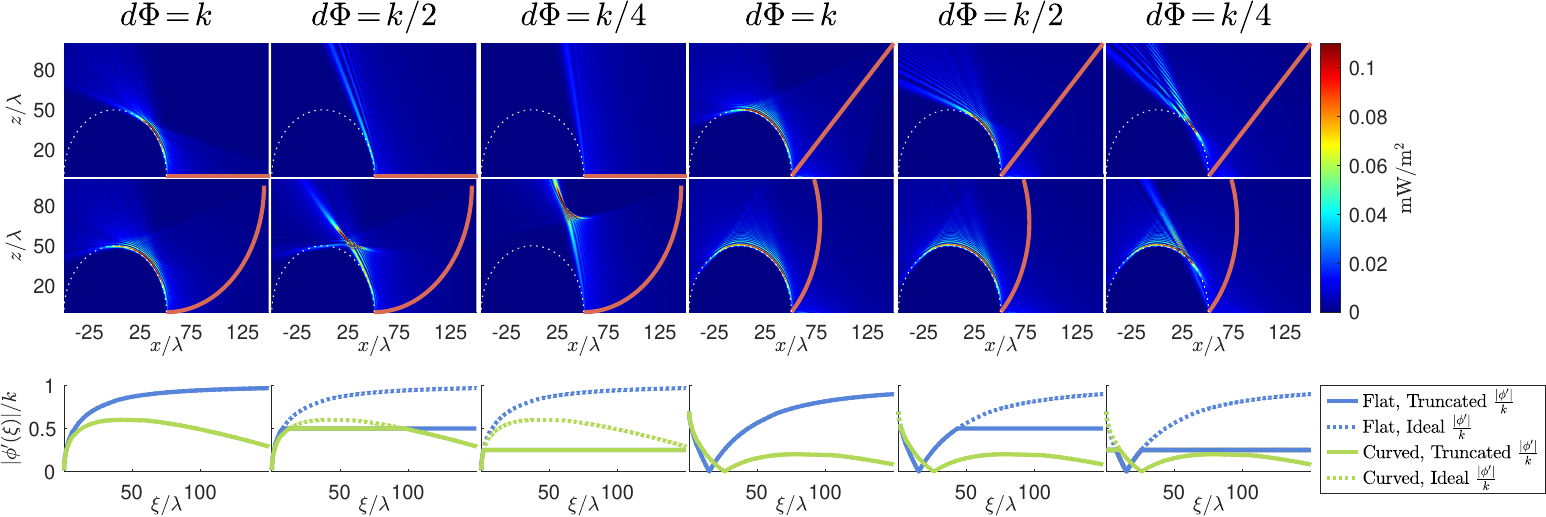}
 \caption{
 Caustic beam with a desired curved trajectory (dashed line) generated with straight and curved apertures (red) with different values of $d\Phi$, together with the ideal phase profile to generate the trajectory and the actual truncated one.
 }
 \label{fig:dPcomparison_dxLambda10}
 \vspace{-0.4cm}
\end{figure*} 

From such considerations, we can claim that trajectories with constant curvature sign (either starting at $\xi(0)\!=\!\xi_m$ or $\xi_M$, or non-intersecting the aperture) can only be generated by a straight aperture if $\Delta_{\vf}\!<\!\pi$. Also, the maximum phase variation $|\phi'|_M$ is at least $|\phi'|_M \!=\! k|\cos(\frac{\pi\!+\!\Delta_{\vf}}{2})|$. 

\subsection{Curved Aperture}
Finally, we consider curved apertures. In particular, we show that the shape of the aperture $\sigmab(\xi)$ can be tailored to the trajectory to minimize the value of $|\phi'|_M$, i.e., $|\phi'|\equiv0$. From~\eqref{eq:phase_profile_free}, the condition for maintaining a constant phase gradient, i.e., $|\phi'|\equiv0$, requires that the tangent to the aperture at any given point is orthogonal to the tangent vector at the corresponding intersection point along the trajectory, i.e.,
\begin{equation}
\phi'(\xi) \!=\! k\hat{\vf'}(t_c(\xi))^T\sigmab'(\xi) \!=\! 0\Leftrightarrow\hat{\vf'}(t_c(\xi))^T\sigmab'(\xi) \!=\! 0. \label{eq:condition1}
\end{equation}
where $t_c(\xi)$ is found by solving
\begin{equation}
 \vf(t_c(\xi))\!-\!\sigmab(\xi)\!=\! \hat{\vf}'(t_c(\xi))\|\vf(t_c(\xi))\!-\!\sigmab(\xi)\|. \label{eq:condition2}
\end{equation}

Moreover, by imposing the condition $\xi(t_c) \!=\! t$, where $t$ represents the arc-length parameter of $\vf$, and substituting this condition in \eqref{eq:condition2} while simultaneously satisfying \eqref{eq:condition1}, one obtains a family of apertures described by
\begin{equation}
 \sigmab(t)\!=\!\vf(t)\!-\!(t\!+\!a)\hat{\vf'}(t),\,\,\forall a\in\mathbb{R}, \,\, a\!\geq\!0. \label{eq:optimal_curve}
\end{equation}
Thus, we can claim that an aperture yielding $|\phi'|\equiv0$ \textit{always exists} and belongs to the family of $\sigmab(\xi)$ in \eqref{eq:optimal_curve}.

\section{Results}
\label{sec:results}
We consider both straight and curved apertures. The apertures are sampled with $d\xi\!=\!\lambda/10$, while the field is evaluated with a granularity of $dx\!=\!dz\!=\!\lambda/20 \!=\! 10^{-3}$m. The working frequency is $15$ GHz ($\lambda\!=\!2$cm). We aim to generate trajectories defined as $\vf(t)\!=\!R(\cos(\frac{t}{R}),\sin(\frac{t}{R}))^{\tran}$, with $t\in[0,\pi R]$, and $R=50\lambda$. The phase profile $\phi(\xi)$ and the field $\psi(x,z)$ are found by numerical integration of \eqref{eq:phase_profile_free_int}, and \eqref{eq:curvedIntegral}, respectively. $t_c(\xi)$ is obtained from \eqref{eq:condition2} either in a closed-form expressions or numerically, depending on $\vf(t)$ and $\sigmab(\xi)$. Straight apertures are defined as $\sigmab(\xi)\!=\!\sigmab_0\! + \!\xi \left(\cos(\theta), \sin(\theta)\right)^{\tran}$, and curved apertures defined as $\sigmab(\xi)\!=\!\sigmab_0 + S(\sin(\frac{\xi}{S}+\theta)\!-\!\sin(\theta), \cos(\theta)\!-\!\cos(\frac{\xi}{S}+\theta))^{\tran}$, with $\sigmab_0 = (R,0)^{\tran}$, $S=\frac{300\lambda}{\pi}$, $\xi\in[0,150\lambda]$, and $\theta=\{0,\pi/4\}$.
The field amplitude at the aperture is set to $A(\xi)\!=\!A_0\frac{1}{1\!+\!\xi}$, to have a uniform intensity distribution on the trajectory \cite{trajectories}. $A_0$ is a scaling factor that bounds the transmission power over the aperture to $1$ W.
From \eqref{eq:beta_angle_progression}, in the case of a flat apertures, we have $\beta(t_c(\xi))\!=\!\frac{\pi}{2}\!-\!\theta\!+\!\frac{t_c(\xi)}{R}$ with $t_c(\xi) \in t(\mathcal{D}) \supseteq [0,\frac{\pi}{2}R]$. Which, from \eqref{eq:general_case_flat}, corresponds to $|\phi'|_M \!=\! k$ in the $\theta=0$ flat aperture.

Fig.~\ref{fig:dPcomparison_dxLambda10} compares the power density $\frac{|\phi(\vx)|^2}{2Z_0}$ ($Z_0 \!=\!377\Omega$) of the generated field with straight and curved apertures for different values of $d\Phi\!=\!\{k,k/2, k/4\}$. 
Focusing on the straight apertures (first row), we can see that,  assuming a non-stringent $d\Phi$, i.e., $d\Phi \!\geq\! |\phi'|_{M}$, the field follows the desired caustic curve without perturbations. Differently, when introducing a more stringent $d\Phi$ for the phase profile, we observe perturbations and losses. Indeed, when enforcing $d\Phi\!=\!k/2$ and $k/4$, a reduction of the power density of the field is visible in the latter portion of the trajectory (lower abscissa values). This is due to the rays originating from the right-hand side of the aperture, which do not contribute constructively to the caustic as $|\phi'(\xi)|\!>\!d\Phi$ in those points. Nonetheless, it is also visible that the tilt of the aperture ($\theta$) helps mitigate field perturbations and cover a larger portion of the trajectory. Focusing on the curved apertures (second row), we observe that they are more resilient to stringent values of $d\Phi$ and are generally capable of generating beams that cover a larger portion of the target caustic, with a higher power density.

\begin{figure}
\centering
\includegraphics[width=0.24\textwidth]{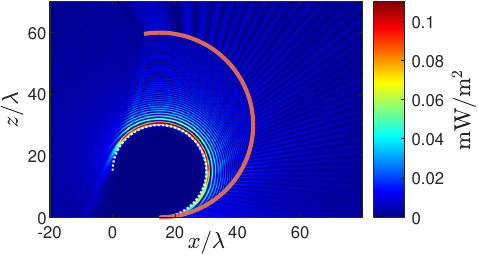}
\caption{
Example of a caustic beam targeting a sharply curved trajectory generated with an aperture tailored to the trajectory.}
\label{fig:sharp_curve}\vspace{-5mm}
\end{figure}

Finally, we show in Fig.~\ref{fig:sharp_curve} an additional example of a curved aperture, tailored to a sharper caustic using \eqref{eq:optimal_curve}. In particular, we consider $\vf(t)\!=\!R(\cos(\frac{t}{R})^{\tran},\sin(\frac{t}{R}))\!+\!(R,R)$, with $t\in[-\frac{\pi}{2}R,0.95\pi R]$ and $R\!=\!15\lambda$. Notably, we can see that the aperture can generate the desired field.  Also, this caustic could not be generated with a straight aperture due to the condition $\Delta_{\vf}\!\geq\!\pi$, as shown in Section~\ref{sec:limits}. These results also suggest that fixed apertures may be suboptimal for certain caustics, motivating future work on reconfigurable aperture designs.
Results show that curved apertures can surpass the limitations of flat apertures, both in phase control requirements and in engineering caustic trajectories.

\section{Conclusions}
\label{sec:conclusion}
In this letter, we have explored the use of curved apertures to overcome the inherent limitations of traditional apertures in \acrlong{bte}. Indeed, as demonstrated, flat apertures impose more stringent constraints on the control of the phase over the aperture itself compared to curved ones. Thanks to the proposed formulation, we can flexibly engineer customized trajectories on arbitrarily shaped apertures, thereby offering a robust and flexible framework for caustic beam shaping. Theoretical analysis and numerical simulations confirmed that curved apertures enable continuous, complex beam paths without external phase-correction mechanisms. Additionally, curved apertures could provide superior phase modulation control, generating smooth wave trajectories with relaxed constraints on the required phase fluctuations.

\bibliographystyle{IEEEtran}
\bibliography{bibliography}

\end{document}

%% file: definitions.tex
\makeatletter
\let\oldabs\abs
\def\abs{\@ifstar{\oldabs}{\oldabs*}}

\let\oldnorm\norm
\def\norm{\@ifstar{\oldnorm}{\oldnorm*}}
\makeatother
\let\vv\relax

\DeclareMathOperator{\vf}{\mathbf{f}}

\DeclareMathOperator{\vk}{\mathbf{k}}

\DeclareMathOperator{\vn}{\mathbf{n}}

\DeclareMathOperator{\vr}{\mathbf{r}}

\DeclareMathOperator{\vv}{\mathbf{v}}

\DeclareMathOperator{\vx}{\mathbf{x}}
\DeclareMathOperator{\vy}{\mathbf{y}}

\DeclareMathOperator{\mE}{\mathbf{E}}

\DeclareMathOperator{\sigmab}{\boldsymbol{\sigma}}

\DeclareMathOperator{\Real}{\mbox{$\mathbb{R}$}}

\DeclareMathOperator{\argmin}{\operatornamewithlimits{argmin}}
\DeclareMathOperator{\argmax}{\operatornamewithlimits{argmax}}

\DeclareMathOperator{\tran}{\mathrm{T}}


%% file: Acro.tex
\newacronym{2d}{2D}{two dimensional}
\newacronym{ris}{RIS}{reconfigurable intelligent surface}
\newacronym{mvt}{MVT}{mean value theorem}
\newacronym{dft}{DFT}{discrete Fourier transform}
\newacronym{idft}{IDFT}{inverse discrete Fourier transform}
\newacronym{ft}{FT}{Fourier transform}
\newacronym{ift}{IFT}{inverse Fourier transform}
\newacronym{fdtd}{FDTD}{Finite-Difference Time-Domain}
\newacronym{fdfd}{FDFD}{Finite-Differences Frequency-Domain}
\newacronym{em}{EM}{electromagnetic}
\newacronym{bte}{BTE}{beam trajectory engineering}